\documentclass[aps,pra,twocolumn,showpacs,floatfix]{revtex4}
\usepackage{epsfig}
\usepackage{graphicx}
\usepackage{dcolumn}
\usepackage{longtable}
\usepackage{amsthm,amsmath}

\begin{document}
%\preprint{\today}

\title{Transition properties of potassium atom}
\vspace{0.5cm}

\author{D. K. Nandy$^1$ \footnote{Email: dillip@prl.res.in}, Yashpal Singh$^1$, B. P. Shah$^2$ and B. K. Sahoo$^1$ \footnote{Email: bijaya@prl.res.in}}
\affiliation{$^1$Theoretical Physics Division, Physical Research Laboratory, Ahmedabad-380009, India\\
$^2$Department of Physics, Faculty of Science, The M. S. University of Baroda, Vadodara-390012, India}

\date{Received date; Accepted date}

\vskip1.0cm

\begin{abstract}
We report here oscillator strengths, transition rates, branching ratios and lifetimes 
due to allowed transitions in potassium (K) atom. We evaluate electric dipole (E1)
amplitudes using an all order relativistic many-body perturbation method. The
obtained results are compared with previously available experimental and theoretical
studies. Using the E1 matrix elements mentioned above and estimated from the 
lifetimes of the $4P$ states, we determine precise values of static
and dynamic polarizabilities for the first five low-lying states in the considered
atom. The static polarizabilities of the ground and $4P$ states in the present work
are more precise than the available measurements in these states. Only the present
work employs relativistic theory to evaluate polarizabilities in the $3D$ states
for which no experimental results are known to compare with. We also reexamine
"magic wavelengths" for the $4P_{1/2} \rightarrow 4S$ and $4P_{3/2} \rightarrow 4S$
transitions due to the linearly polarized light which are useful to perform 
state-insensitive trapping of K atoms.
\end{abstract} 

\pacs{}
\maketitle

\section{Introduction}
Potassium (K) atom is one of the suitable alkali atoms whose two stable isotopes 
are fermions and one of them is boson that makes it a special system for cold
atom studies \cite{wu,catani}. Studies of its transition properties are very useful 
in manipulating trapping and cooling of this atom \cite{allen,zwierlein,mckay}. Especially
to find out its "magic wavelengths" at which the differential Stark shifts of a 
transition become zero \cite{katori}, accurate knowledge of transition properties are necessary 
\cite{arora1,arora2}. Sophisticated experiments are generally performed to measure these magic
wavelengths \cite{mejri,barber}, however they can also be determined precisely by evaluating differential
polarizabilities using accurate relativistic many-body methods \cite{arora1,arora2}. In fact, K atom
is an one valence electron system, whose properties can be evaluated precisely that
can ultimately act as benchmark tests for the experimental results.
Also, a suitable trapping technique of alkali atoms is very much useful in various fields like
in the atomic clock, quantum computing, quantum information etc. experiments. In an optical lattice atomic clock
experiment, the atoms are trapped by a periodic potential formed by the applied laser beam 
\cite{takamoto,derevianko} for which the trapped atoms interact with the oscillating electric field of the trapping beam that causes
the shift in the energy levels inside the atoms. But in order to achieve maximum stability of an atomic clock, 
it is very crucial to tune the laser beam at a particular wavelength such that there will be no effects
to the internal clock transition frequency \cite{katori,barber,derevianko}. These effects can be
estimated by estimating dynamic polarizabilities of the atomic states \cite{katori,arora1,arora2,barber,derevianko}.

Accurate values of transition properties in K are also useful in the astrophysical
studies \cite{burrows, zeippen, charro}. K is produced in massive stars after oxygen burning and
its isotopes abundances are analyzed in 58 metal-poor stars. In these analysis, theoretical
estimate of ratios of abundances of [K/Fe] with [Fe/H] during the chemical evolution of Galaxy 
are compared with the observations \cite{Zhang}.
There are also various astrophysical objects such as L dwarfs, T dwarfs and irradiated giant planets
whose spectra are dominated by absorption lines of many alkali atoms \cite{burgasser,burrows2,
burrows3,charb}. Many telescopes operating at various wavelengths from ultraviolet (UV) to infrared
(IR) range are used to study various spectral properties of astronomical objects, but these IR 
telescopes require information about detailed spectroscopic data of atomic systems such as
oscillator strengths, transition probabilities etc. \cite{johansson, jorissen, pickering}. Also, 
identification of new IR lines will be useful in understanding equilibrium temperature, gravity and 
abundance for many ultra-cool dwarf stars as well as extra solar giant planets in our Galaxy 
\cite{lyubchik}. Therefore it is always very useful to have enough atomic information on new IR 
atomic lines based on high precision relativistic calculations, so that they can be used in the analysis
of spectral measurements generally carried out by sophisticated IR telescopes.

In this paper, we  evaluate electric dipole (E1) matrix elements and determine oscillator strengths,
transition probabilities, branching ratios, lifetimes and polarizabilities of many states of K atom.
Furthermore, we determine the magic wavelengths due to the linearly polarized light for the $4P_{1/2}
\rightarrow 4S$ and $4P_{3/2} \rightarrow 4S$ transitions and compare them with the available results. 

The rest of the paper is organized as follows: in Sections II and III, we give the theories
of transition probability due to allowed transition and linear Stark shift and describe briefly
about the method of calculation. In the next section, we present the results and discuss 
about them. Unless stated otherwise, we use atomic unit (a.u.) for all the physical properties 
through out the paper.

\section{Theory and Method of Calculations}
\subsection{Allowed radiative properties}
The transition probability (in $s^{-1}$) due to E1 matrix element from atomic state $|\Psi_k \rangle$
to $|\Psi_i \rangle$ is given by
\begin{eqnarray}
A_{ki}= \frac{2.0261\times 10^{-6}}{\lambda_{ki}^3 g_k} S_{ki} \label{eqn1},
\end{eqnarray}
where $S_{ki}= |\langle \Psi_k ||D|| \Psi_i\rangle|^2$ is the line strength, $\lambda_{ki}$ is
the wavelength (in $cm^{-1}$) and $g_k=2J_k+1$ is the degeneracy factor for angular momentum 
$J_k$ of the state $|\Psi_k \rangle$.

The relative intensities of radiative transitions are generally estimated by their oscillator
strengths that for an allowed transition is given by
\begin{eqnarray}
f_{ki} = 1.4992\times 10^{-24} A_{ki} \frac{g_k}{g_i}\lambda_{ki}^2,
\label{eqn2}
\end{eqnarray}
which again for the emission and absorption lines are related as
\begin{eqnarray}
 g_i f_{ik} = - g_k f_{ki}.
\label{eqn3}
\end{eqnarray}

The lifetime of the $|\Psi_k\rangle$ state due to the allowed transitions (transition probabilities
via other multi-poles are neglected for the considered states in the present atom) is
estimated by
\begin{eqnarray}
\tau_k &=& \frac {1} {\sum_{i} A_{ki}}.
\label{eqn4}
\end{eqnarray}
The branching ratio of a transition probability from the state $|\Psi_k\rangle$ 
to a lower state $|\Psi_i\rangle$ is given by
\begin{eqnarray}
\Gamma_{ki} &=& \frac {A_{ki}} {\sum_{l} A_{kl}} \nonumber \\ 
 &=& \tau_k A_{k i},
\label{eqn5}
\end{eqnarray}
where sum over $l$ represents all possible allowed transitions from $|\Psi_k\rangle$.

\subsection{Linear Stark shift}
In the application of external electric field described by a plane electromagnetic wave of angular frequency $\omega$, 
the energy level of a state $|\Psi_k\rangle$ of an atom shifted by
\begin{eqnarray}
\Delta E_k(\omega) &=& -\frac{1}{2}\alpha_k(\omega) \mathcal{E}^2,
\label{seqn}
\end{eqnarray}
where $\alpha_k$ is the known as polarizability of the $|\Psi_k \rangle$ state and $\mathcal{E}$ is the strength
of the applied electric field. In a more general form $\alpha_k$ in $M_{J_k}$ quantum number independent parameters is expressed as
\begin{eqnarray}
\alpha_k(\omega)=\alpha^0_v(\omega)+\frac{3M^2_{J_k}-J_k(J_k+1)}{J_k(2J_k-1)}\alpha^2_k(\omega), 
\end{eqnarray}
with the parameters $\alpha^0_k(\omega)$ and $\alpha^2_k(\omega)$ 
are called as scalar and tensor polarizabilities, respectively. They are given in terms of the reduced
matrix elements of dipole operator $D$ as
\begin{eqnarray}
\alpha_k^{0}(\omega)&=& \frac{1}{{3(2J_k+1)} }\sum_l 
|\langle J_k \parallel D \parallel J_l \rangle|^{2} \nonumber \\
&\times& \left [ \frac{1}{E_l-E_k+\omega}+\frac{1}{E_l-E_k-\omega} \right ] 
\end{eqnarray}
and
\begin{eqnarray}
\alpha_k^{2}(\omega) &=& -2 \sqrt { \frac{5J_k(2J_k-1)} {6(J_k+1)(2J_k+1)(2J_k+3)} } \nonumber \\
&\times& \sum_l \left \{ \begin{array}{ccc} J_k & 2 & J_k \\ 1 & J_l & 1 \end{array} \right \}
(-1)^{J_k+J_l+1} 
|\langle J_k \parallel D \parallel J_l \rangle|^{2} \nonumber \\
&\times& \left [ \frac{1}{E_l-E_k+\omega}+\frac{1}{E_l-E_k-\omega} \right ] ,
\end{eqnarray}
for all the allowed states $|\Psi_l\rangle$ from $|\Psi_k\rangle$.

Magic wavelengths of a transition can be determined by evaluating Stark shifts for 
both the associated states and finding out $\omega$ values where the shifts are equal for
these states. Owing to the fact that these effects are directly proportional to the differential
polarizabilities of the associated states, magic wavelengths are finally correspond to finding 
values of $\omega$ that produces null differential dynamic polarizabilities of the transition.

\subsection{Method of calculations}
In the coupled-cluster ansatz, we express the atomic states with a closed core and one valence
orbital as
\begin{eqnarray}
| \Psi_k \rangle &=& e^T \{1+S_k\} | \Phi_k \rangle , 
\label{eqn10}
\end{eqnarray}
where initial state is constructed as $\vert \Phi_k \rangle = a_k^{\dagger} |\Phi_0 \rangle$ with 
$|\Phi_0\rangle$ is the mean-field wave function for the closed core obtained by Dirac-Hartree-Fock
(DHF) method and $a_k^{\dagger}$ represents appending the valence orbital denoted by $k$. Here
$T$ and $S_k$ are the excitation operators that account correlation effects to all orders by exciting
electrons from core orbitals and valence along with core orbitals from the corresponding DHF states,
respectively. In the present calculations, only singly and doubly excited configurations are generated using
both the $T$ and $S_k$ operators in the CCSD method framework; also, important correlation effects involving
valence electron are accounted through the $S_k$ operator perturbatively in the CCSD(T) method 
approximation. Discussions on these approaches can be found more elaborately in \cite{bartlett}.
The details of single particle construction and active space for the calculation of
atomic wave functions for the considered atom are described in our previous work \cite{yashpal}.

Following the approach given in \cite{arora}, we evaluate polarizabilities of the considered states
in K atom by expressing different contributions as
\begin{eqnarray}
\alpha^{\lambda}_k= \alpha^{\lambda}_k(c)+\alpha^{\lambda}_k(vc)+\alpha^{\lambda}_k(v)
\end{eqnarray}
where superscript $\lambda$ with values 0 and 1 correspond to the scalar and tensor polarizabilities,
respectively, and notations $c$, $vc$ and $v$ in the parentheses represent correlation contributions 
from core, core-valence and valence orbitals, respectively. It has to be noted that the core contribution to 
the tensor polarizability is zero.

\begin{table}[t]
\caption{Results of attachment energies (EAs) in $cm^{-1}$ of different states K using
DHF, CCSD and CCSD(T) methods. The results are compared with the experimental values 
given in NIST data table \cite{Nist} and difference between NIST data and results 
obtained from the CCSD(T) method are quoted as $\Delta$ in percentage (\%).}
\begin{ruledtabular}
\begin{center}
\begin{tabular}{lcccccc}
State    &  DHF & CCSD & CCSD(T) & NIST \cite{Nist} & $\Delta$\\
\hline
       &  & & \\
$4s \ ^2S_{1/2}$ &$-32370.47 $&$-35016.98$  & $- 35016.59$ &$-35009$ & 0.02 \\

$4p \ ^2P_{1/2}$ &$-21006.47 $&$-22010.66$  & $- 22013.81$ &$-22025$ & 0.05 \\
$4p \ ^2P_{3/2}$ &$-20959.41 $&$-21950.59$  & $- 21953.72$ &$-21967$ & 0.06 \\

$5s \ ^2S_{1/2}$ &$-13407.12 $&$-13982.93$  & $- 13982.35$ &$-13983$ & $0.004$ \\

$3d \ ^2D_{5/2}$ &$-12747.04 $&$-13418.72$  & $- 13417.33$ &$-13475$ & 0.44 \\
$3d \ ^2D_{3/2}$ &$-12744.31 $&$-13417.27$  & $- 13415.78$ &$-13473$ & 0.41 \\

$5p \ ^2P_{1/2}$ &$-10012.08 $&$-10303.07$  & $- 10304.42$ &$-10308$ & 0.03 \\
$5p \ ^2P_{3/2}$ &$-9995.88  $&$-10283.56$  & $- 10284.89$ &$-10290$ & 0.05 \\

$4d \ ^2D_{5/2}$ &$-7206.85  $&$-7572.74$   & $- 7571.24$  &$-7613$  & 0.53 \\
$4d \ ^2D_{3/2}$ &$-7205.25  $&$-7571.88$   & $- 7570.84$  &$-7612$  & 0.53 \\

$6s \ ^2S_{1/2}$ &$-7338.12  $&$-7558.71$   & $- 7558.34$  &$-7559$  & 0.01 \\

$4f \ ^2F_{5/2}$ &$-6859.21  $&$-6881.92$   & $- 6881.95$  &$-6882$  & 0.00 \\
$4f \ ^2F_{7/2}$ &$-6859.21  $&$-6881.92$   & $- 6881.94$  &$-6882$  & 0.00 \\

$6p \ ^2P_{1/2}$ &$-5881.48  $&$-6008.02$   & $- 6008.75$  &$-6011$  & 0.04  \\
$6p \ ^2P_{3/2}$ &$-5874.06  $&$-5999.25$   & $- 5999.98$  &$-6002$  & 0.03 \\

$5d \ ^2D_{5/2}$ &$-4596.71  $&$-4800.51$   & $- 4799.29$  &$-4825$  & 0.51 \\         
$5d \ ^2D_{3/2}$ &$-4595.80  $&$-4800.46$   & $- 4799.23$  &$-4824$  & 0.51  \\

$7s \ ^2S_{1/2}$ &$-4627.27  $&$-4735.10$   & $- 4734.85$  &$-4736$  & 0.02 \\

$5f \ ^2F_{7/2}$ &$-4390.01  $&$-4403.00$   & $- 4403.01$  &$-4403$  & 0.00 \\
$5f \ ^2F_{5/2}$ &$-4390.01  $&$-4403.00$   & $- 4403.01$  &$-4403$  & 0.00 \\

$5g \ ^2G_{7/2}$ &$-4389.50  $&$-4390.90$   & $- 4390.91$  &$-4392$  & 0.02 \\
$5g \ ^2G_{9/2}$ &$-4389.50  $&$-4390.90$   & $- 4390.90$  &$-4392$  & 0.02 \\

$7p \ ^2P_{1/2}$ &$-3871.88  $&$-3938.52$   & $- 3938.99$  &$-3940$  & 0.02 \\
$7p \ ^2P_{3/2}$ &$-3867.88  $&$-3933.84$   & $- 3934.31$  &$-3935$  & 0.02 \\

$6d \ ^2D_{5/2}$ &$-3177.17  $&$-3298.89$   & $- 3298.31$  &$-3314$  & 0.47 \\
$6d \ ^2D_{3/2}$ &$-3176.62  $&$-3298.69$   & $- 3297.97$  &$-3314$  & 0.48 \\

$8s \ ^2S_{1/2}$ &$-3182.99  $&$-3244.03$   & $- 3243.85$  &$-3245$  & 0.03 \\
\end{tabular}
\end{center}
\end{ruledtabular}
\label{tab1}
\end{table}
\section{Results and Discussion}
We present below first the electron attachment energies (EAs) of valence electrons from the states 
having different valence orbitals with the closed core $[3p^6]$. Then we present E1 matrix elements
and using them with the experimental energies, we determine transition probabilities
and other related transition properties. Also, we evaluate $\alpha^{\lambda}_k(v)$ for various states from
these E1 matrix elements and experimental energies. Other $\alpha^{\lambda}_k(c)$ and 
$\alpha^{\lambda}_k(vc)$ contributions to $\alpha_k$, which are found to be smaller in magnitudes
compared to $\alpha^{\lambda}_k(v)$, are evaluated using a third order many-body perturbation theory
(MBPT(3) method) following similar approaches as employed in \cite{sahoo1,sahoo2,sahoo3}. Using the dynamic 
polarizabilities, we verify the "magic wavelengths" in the $4P_{1/2} \rightarrow 4S$ and
$4P_{3/2} \rightarrow 4S$ transitions and compare them with previously reported values.

%\begin{center}
\begin{longtable}{lllll}
\caption{Calculated E1 matrix elements and their line strengths in a.u. for all the allowed transitions 
among the states up to $8S$ states.} \label{tab2}\\
\hline \hline  
\multicolumn{2}{c}{Transition $k \rightarrow i$} & DHF & CCSD(T) & $S_{k i}$ \\ \hline \\
\endfirsthead

\multicolumn{4}{r}{\tablename\ \thetable{} -- continued from previous page.} \\
\hline \hline
\multicolumn{2}{c}{Transition $k \rightarrow i$} & DHF & CCSD(T) & $S_{k i}$ \\ \hline \\
\endhead

\hline
\multicolumn{4}{r}{ {\it Continue} \dots } \\
\endfoot

\hline \hline
\endlastfoot
$4p \ ^2P_{1/2}$ & $\rightarrow 4s \ ^2S_{1/2}$  &$4.554$   &$4.131(20)$       &17.065(160)               \\
                                                                              
$4p \ ^2P_{3/2}$ & $\rightarrow 4s \ ^2S_{1/2}$  &$6.439$   &$5.841(20)$       &34.117(234)           \\
                 
$5s \ ^2S_{1/2}$ & $\rightarrow 4p \ ^2P_{3/2}$  &$5.658$   &$5.524(20)$       &$30.515(221)$           \\
                 & $\rightarrow 4p \ ^2P_{1/2}$  &3.974     &3.876(10)         &15.023(77)          \\
                                                                              
$3d \ ^2D_{5/2}$ & $\rightarrow 4p \ ^2P_{3/2}$  &$11.564$  &$10.749(50)$      &$115.54(107)$         \\

$3d \ ^2D_{3/2}$ & $\rightarrow 4p \ ^2P_{3/2}$  &$3.855$   &$3.583(20)$       &$12.838(143)$           \\
                 & $\rightarrow 4p \ ^2P_{1/2}$  &$8.596$   &$7.988(40)$       &$63.808(639)$                \\

$5p \ ^2P_{1/2}$ & $\rightarrow 3d \ ^2D_{3/2}$  &$8.198$   &$7.278(130)$      &$52.97(189)$   \\
                 & $\rightarrow 5s \ ^2S_{3/2}$  &$9.935$   &$9.489(10)$       &$90.041(190)$                    \\
                 & $\rightarrow 4s \ ^2S_{5/2}$  &$0.312$   &$0.282(6)$        &$0.079(0.003)$                     \\

$5p \ ^2P_{3/2}$ & $\rightarrow 3d \ ^2D_{5/2}$  &$10.955$  &$9.729(150)$      &$94.65(292)$                    \\
                 & $\rightarrow 3d \ ^2D_{3/2}$  &3.655     &$3.242(50)$       &$10.511(324)$                        \\   
                 & $\rightarrow 5s \ ^2S_{1/2}$  &$14.031$  &$13.399(20)$      &$179.533(536)$          \\
                 & $\rightarrow 4s \ ^2S_{1/2}$  &$0.456 $  &$0.416(6)$        &$0.173(5)$          \\

$4d \ ^2D_{5/2}$ & $\rightarrow 5p \ ^2P_{3/2}$  &$23.116$  &$22.842(300)$     &$521.8(137)$               \\  
                 & $\rightarrow 4p \ ^2P_{3/2}$  &$1.003 $  &$0.260(5)$        &$0.067(3)$                   \\

$4d \ ^2D_{3/2}$ & $\rightarrow 5p \ ^2P_{3/2}$  &7.704      &$7.613(100)$     &$57.96(152)$                \\ 
                 & $\rightarrow 5p \ ^2P_{1/2}$  &17.173     &$16.969(240)$    &$287.95(814)$                   \\  
                 & $\rightarrow 4p \ ^2P_{3/2}$  &0.336      &$0.088(5)$       &$0.008(1)$                      \\
                 & $\rightarrow 4p \ ^2P_{1/2}$  &0.769      &$0.220(5)$       &$0.048(2)$     \\

$6s \ ^2S_{1/2}$ & $\rightarrow 5p \ ^2P_{3/2}$  &12.656     &$12.470(20)$     &$155.50(50)$                 \\
                 & $\rightarrow 5p \ ^2P_{1/2}$  &$8.898$    &$8.760(10)$      &$76.738(175)$                \\
                 & $\rightarrow 4p \ ^2P_{3/2}$  &1.309      &$1.287(10)$      &$1.656(26)$           \\
                 & $\rightarrow 4p \ ^2P_{1/2}$  &$0.925$    &$0.909(10)$      &$0.826(18)$               \\

$4f \ ^2F_{5/2}$ & $\rightarrow 4d \ ^2D_{3/2}$  &$25.230$   &$25.37(146)$     &$643.8(741)$        \\
                 & $\rightarrow 4d \ ^2D_{5/2}$  &6.743      &$6.78(38) $      &$46.0(52)$           \\
                 & $\rightarrow 3d \ ^2D_{3/2}$  &$14.112$   &$12.41(11)$      &$154.1(27)$             \\
                 & $\rightarrow 3d \ ^2D_{5/2}$  &$3.769 $   &$3.318(30)$      &$11.01(20)$               \\

$4f \ ^2F_{7/2}$ & $\rightarrow 4d \ ^2D_{5/2}$  &30.158     &$30.34(170)$     &$923(103)$   \\
                 & $\rightarrow 3d \ ^2D_{5/2}$  &$16.857$   &$14.84(12)$      &$220.3(36)$                    \\

$6p \ ^2P_{1/2}$ & $\rightarrow 6s \ ^2S_{1/2}$  &17.195     &$16.613(20)$     &$275.99(66)$                    \\
                 & $\rightarrow 4d \ ^2D_{3/2}$  &16.833     &$14.76(37)$      &$217.9(109)$                        \\   
                 & $\rightarrow 3d \ ^2D_{3/2}$  &$1.025$    &$1.037(10)$      &$1.075(21)$          \\
                 & $\rightarrow 5s \ ^2S_{1/2}$  &$0.873$    &$0.906(10)$      &$0.820(18)$          \\  
                 & $\rightarrow 4s \ ^2S_{1/2}$  &$0.104$    &$0.087(5)$       &$0.008(1)$               \\

$6p \ ^2P_{3/2}$ & $\rightarrow 6s \ ^2S_{1/2}$  &24.273     &$23.444(20)$     &$549.621(938)$               \\  
                 & $\rightarrow 4d \ ^2D_{3/2}$  &$7.502 $   &$6.57(16)$       &$43.20(210) $                  \\    
                 & $\rightarrow 4d \ ^2D_{5/2}$  &$22.487$   &$19.734(470)$    &$389.4(185)$                  \\    
                 & $\rightarrow 3d \ ^2D_{3/2}$  &0.462      &$0.467(3)$       &$0.218(3)$                   \\
                 & $\rightarrow 3d \ ^2D_{5/2}$  &$1.387 $   &$1.389(10)$      &$1.954(28)$                  \\
                 & $\rightarrow 5s \ ^2S_{1/2}$  &$1.263 $   &$1.312(6)$       &$1.721(15)$      \\
                 & $\rightarrow 4s \ ^2S_{1/2}$  &$0.155 $   &$0.132(6)$       &$0.017(2)$                  \\

$5d \ ^2D_{5/2}$ & $\rightarrow 6p \ ^2P_{3/2}$  &38.150     &$38.72(74)$      &$1499(57)$          \\
                 & $\rightarrow 4f \ ^2F_{7/2}$  &$5.803$    &$8.42(10)$       &$70.90(17)$               \\
                 & $\rightarrow 4f \ ^2F_{5/2}$  &1.297      &$1.883(20)$      &$3.546(75)$              \\
                 & $\rightarrow 5p \ ^2P_{3/2}$  &2.690      &$1.461(5)$       &$2.134(15)$                \\
                 & $\rightarrow 4p \ ^2P_{3/2}$  &0.122      &$0.374(5)$       &$0.141(4)$            \\

$5d \ ^2D_{3/2}$ & $\rightarrow 6p \ ^2P_{3/2}$  &12.712     &$12.90(26)$      &$166.5(67)$    \\
                 & $\rightarrow 6p \ ^2P_{1/2}$  &28.333     &$28.76(57)$      &$827.3(328)$     \\
                 & $\rightarrow 4f \ ^2F_{5/2}$  &$4.844$    &$7.034(90)$      &$49.48(127)$    \\
                 & $\rightarrow 5p \ ^2P_{3/2}$  &0.899      &$0.490(5)$       &$0.240(5)$               \\
                 & $\rightarrow 5p \ ^2P_{1/2}$  &2.042      &$1.138(10)$      &$1.295(23)$                  \\
                 & $\rightarrow 4p \ ^2P_{3/2}$  &0.042      &$0.124(5)$       &$0.015(1)$                    \\  
                 & $\rightarrow 4p \ ^2P_{1/2}$  &0.105      &$0.264(5)$       &$0.071(3)$                 \\
 
$7s \ ^2S_{1/2}$ & $\rightarrow 6p \ ^2P_{3/2}$  &22.032     &$21.828(20)$     &$476.46(87)$    \\
                 & $\rightarrow 6p \ ^2P_{1/2}$  &$15.496$   &$15.342(10)$     &$235.38(31)$     \\
                 & $\rightarrow 5p \ ^2P_{3/2}$  &2.624      &$2.563(10) $     &$6.569(51)$    \\
                 & $\rightarrow 5p \ ^2P_{1/2}$  &$1.857 $   &$1.814(10)$      &$3.291(37)$                \\
                 & $\rightarrow 4p \ ^2P_{3/2}$  &0.686      &$0.677(6)$       &$0.458(8)$                   \\
                 & $\rightarrow 4p \ ^2P_{1/2}$  &$0.485 $   &$0.479(5)$       &$0.229(5)$                    \\

$5f \ ^2F_{7/2}$ & $\rightarrow 5d \ ^2D_{5/2}$  &$56.649$  &$56.34(331)$      &$3174(373)$                \\ 
                 & $\rightarrow 4d \ ^2D_{5/2}$  &$20.464$  &$15.84(20)$       &$250.84(63)$                   \\  
                 & $\rightarrow 3d \ ^2D_{5/2}$  &$6.131 $  &$5.913(40)$       &$34.963(473)$                      \\
 
$5f \ ^2F_{5/2}$ & $\rightarrow 5d \ ^2D_{3/2}$  &$47.394$  &$47.12(289)$      &$2220(272)$                           \\
                 & $\rightarrow 5d \ ^2D_{5/2}$  &$12.667$  &$12.60(74)$       &$158.7(186)$                      \\
                 & $\rightarrow 4d \ ^2D_{3/2}$  &17.142    &$13.24(17)$       &$175.2(45)$                             \\
                 & $\rightarrow 4d \ ^2D_{5/2}$  &$4.576$   &$3.541(40)$       &$12.54(28)$                            \\
                 & $\rightarrow 3d \ ^2D_{3/2}$  &$5.123$   &$4.949(30)$       &$24.49(30)$                          \\
                 & $\rightarrow 3d \ ^2D_{5/2}$  &$1.371$   &$1.322(10)$       &$1.748(26)$                           \\

$5g \ ^2G_{7/2}$ & $\rightarrow 5f \ ^2F_{5/2}$  &$41.688 $  &$41.79(418)$     &$1746(349)$                          \\
                 & $\rightarrow 5f \ ^2F_{7/2}$  &$8.019  $  &$8.04(80)$       &$64.7(128)$                       \\
                 & $\rightarrow 4f \ ^2F_{7/2}$  &$6.313 $   &$6.261(10)$      &$39.20(12)$                       \\
                 & $\rightarrow 4f \ ^2F_{5/2}$  &$32.803$   &$32.53(16)$      &$1058.4(104)$                       \\

$5g \ ^2G_{9/2}$ & $\rightarrow 5f \ ^2F_{7/2}$  &$47.441 $  &$47.58(480)$     &$2264(457)$              \\
                 & $\rightarrow 4f \ ^2F_{7/2}$  &$37.348$   &$37.040(20)$     &$1372(15)$               \\

$7p \ ^2P_{1/2}$ & $\rightarrow 7s \ ^2S_{1/2}$  &$26.351 $  &$25.595(20)$     &$655.1(10)$                  \\
                 & $\rightarrow 5d \ ^2D_{3/2}$  &$27.927 $  &$24.46(68)$      &$598(33)$                   \\
                 & $\rightarrow 6s \ ^2S_{1/2}$  &$1.658  $  &$1.735(10)$      &$3.010(34)$             \\
                 & $\rightarrow 4d \ ^2D_{3/2}$  &$2.313  $  &$2.424(30) $     &$5.876(145)$                  \\
                 & $\rightarrow 3d \ ^2D_{3/2}$  &$0.484  $  &$0.500(6)$       &$0.250(6)$                  \\
                 & $\rightarrow 5s \ ^2S_{1/2}$  &$0.320  $  &$0.341(5)$       &$0.460(3) $                  \\
                 & $\rightarrow 4s \ ^2S_{1/2}$  &$0.053  $  &$0.041(5)$       &$0.0020(4)$                  \\

$7p \ ^2P_{3/2}$ & $\rightarrow 7s \ ^2S_{1/2}$  &$37.187 $   &$36.107(20)$    &$1303.7(14)$                     \\
                 & $\rightarrow 5d \ ^2D_{3/2}$  &$12.445 $   &$10.89(36)$     &$118.50(784)$                  \\
                 & $\rightarrow 5d \ ^2D_{5/2}$  &$37.305 $   &$32.69(90)$     &$1068.5(588)$                    \\
                 & $\rightarrow 6s \ ^2S_{1/2}$  &$2.389  $   &$2.501(10)$     &$6.255(50)$                 \\
                 & $\rightarrow 4d \ ^2D_{3/2}$  &$1.041  $   &$1.089(10)$     &$1.186(22)$                   \\
                 & $\rightarrow 4d \ ^2D_{5/2}$  &$3.126  $   &$3.259(40)$     &$10.62(26)$         \\
                 & $\rightarrow 3d \ ^2D_{3/2}$  &$0.218  $   &$0.225(5)$      &$0.051(2)$                    \\
                 & $\rightarrow 3d \ ^2D_{5/2}$  &$0.656  $   &$0.676(10)$     &$0.587(13)$                   \\
                 & $\rightarrow 5s \ ^2S_{1/2}$  &$0.469  $   &$0.499(5)$      &$0.249(5)$                    \\
                 & $\rightarrow 4s \ ^2S_{1/2}$  &$0.081  $   &$0.064(5)$      &$0.004(1)$                    \\
 
$6d \ ^2D_{5/2}$ & $\rightarrow 7p \ ^2P_{3/2}$  &$56.587$    &$58.30(141)$    &$3399(164)$                     \\
                 & $\rightarrow 5f \ ^2F_{5/2}$  &$2.843 $    &$4.070(60)$     &$16.57(49)$                  \\
                 & $\rightarrow 5f \ ^2F_{7/2}$  &$12.713$    &$18.20(26)$     &$331.39(946)$                    \\
                 & $\rightarrow 6p \ ^2P_{3/2}$  &$4.838 $    &$2.993(20)$     &$8.958(120)$                 \\
                 & $\rightarrow 4f \ ^2F_{5/2}$  &$0.416 $    &$0.541(5)$      &$0.293(5)$                   \\
                 & $\rightarrow 4f \ ^2F_{7/2}$  &$1.862 $    &$2.421(10)$     &$5.861(48)$         \\
                 & $\rightarrow 5p \ ^2P_{3/2}$  &$0.818 $    &$0.045(5)$      &$0.0020(4) $                   \\
                 & $\rightarrow 4p \ ^2P_{3/2}$  &$0.053 $    &$0.404(5)$      &$0.163(4)$                   \\

$6d \ ^2D_{3/2}$ & $\rightarrow 7p \ ^2P_{3/2}$  &$18.854$    &$19.43(48)$     &$377.6(186)$                     \\
                 & $\rightarrow 7p \ ^2P_{1/2}$  &$42.018$    &$43.31(104)$    &$1876(90)$                  \\
                 & $\rightarrow 5f \ ^2F_{5/2}$  &$10.614$    &$15.22(23)$     &$232(7)$                    \\
                 & $\rightarrow 6p \ ^2P_{3/2}$  &$1.617 $    &$1.002(5)$      &$1.004(10)$                 \\
                 & $\rightarrow 6p \ ^2P_{1/2}$  &$3.661 $    &$2.304(20)$     &$5.308(92) $                  \\
                 & $\rightarrow 4f \ ^2F_{5/2}$  &$1.555 $    &$2.029(20)$     &$4.117(81)$         \\
                 & $\rightarrow 5p \ ^2P_{3/2}$  &$0.275 $    &$0.015(5)$      &$0.0002(1)$                    \\
                 & $\rightarrow 5p \ ^2P_{1/2}$  &$0.6346$    &$0.059(5)$      &$0.003(1)$                   \\
                 & $\rightarrow 4p \ ^2D_{3/2}$  &$0.017 $    &$0.135(5)$      &$0.018(1)$                   \\
                 & $\rightarrow 4p \ ^2P_{1/2}$  &$0.031 $    &$0.293(5)$      &$0.086(3)$                     \\

$8s \ ^2S_{1/2}$ & $\rightarrow 7p \ ^2P_{3/2}$  &$33.839$    &$33.63(30)$     &$1130.6(202)$                     \\
                 & $\rightarrow 7p \ ^2P_{1/2}$  &$23.808$    &$23.641(20)$    &$558.90(95)$                  \\
                 & $\rightarrow 6p \ ^2P_{3/2}$  &$4.308 $    &$4.207(6)$      &$17.699(50)$                    \\
                 & $\rightarrow 6p \ ^2P_{1/2}$  &$3.051 $    &$2.982(6)$      &$8.892(36)$                 \\
                 & $\rightarrow 5p \ ^2P_{3/2}$  &$1.331 $    &$1.310(6)$      &$1.716(16)$                   \\
                 & $\rightarrow 5p \ ^2P_{1/2}$  &$0.943 $    &$0.918(6)$      &$0.843(11)$         \\
                 & $\rightarrow 4p \ ^2P_{3/2}$  &$0.452 $    &$0.447(5)$      &$0.211(4)$                    \\
                 & $\rightarrow 4p \ ^2P_{1/2}$  &$0.321 $    &$0.316(5)$      &$0.111(3)$                   \\

$6f \ ^2F_{7/2}$ & $\rightarrow 3d \ ^2D_{5/2}$  &$3.479$     &$3.455(30)$     &$11.94(21)$                     \\

$6f \ ^2F_{5/2}$ & $\rightarrow 3d \ ^2D_{5/2}$  &$0.778$     &$0.777(10)$     &$0.006(16)$                     \\
                 & $\rightarrow 3d \ ^2D_{3/2}$  &$2.910$     &$2.908(30)$     &$8.456(174)$                  \\

$8p \ ^2P_{1/2}$ & $\rightarrow 3d \ ^2D_{3/2}$  &$0.307 $    &$0.321(5)$      &$0.103(3)$                  \\
                 & $\rightarrow 4s \ ^2S_{1/2}$  &$0.033 $    &$0.023(3)$      &$0.0010(1)$                   \\
                
$8p \ ^2P_{3/2}$ & $\rightarrow 3d \ ^2D_{5/2}$  &$0.415 $    &$0.432(5)$      &$0.187(4)$                  \\
                 & $\rightarrow 3d \ ^2D_{3/2}$  &$0.138 $    &$0.144(4)$      &$0.021(1)$                   \\
                 & $\rightarrow 4s \ ^2S_{1/2}$  &$0.051 $    &$0.038(3)$      &$0.0014(2)$             \\

$7d \ ^2D_{5/2}$ & $\rightarrow 4p \ ^2P_{3/2}$  &$0.093$     &$0.356(5)$      &$0.127(4)$                     \\
                 
$7d \ ^2D_{3/2}$ & $\rightarrow 4p \ ^2P_{3/2}$  &$0.031$     &$0.119(3)$      &$0.014(1)$                     \\
                 & $\rightarrow 4p \ ^2P_{1/2}$  &$0.062$     &$0.261(4)$      &$0.068(2)$                  \\
                 
$9s \ ^2S_{1/2}$ & $\rightarrow 4p \ ^2P_{3/2}$  &$0.322$     &$0.317(5)$      &$0.100(3)$                     \\
                 & $\rightarrow 4p \ ^2P_{1/2}$  &$0.228$     &$0.225(3)$      &$0.051(1)$                  \\

$7f \ ^2F_{7/2}$ & $\rightarrow 3d \ ^3D_{5/2}$  &$2.347$     &$2.386(20)$     &$5.693(95)$                     \\

$7f \ ^2F_{5/2}$ & $\rightarrow 3d \ ^3D_{5/2}$  &$0.526$     &$0.535(10)$     &$0.286(10)$                     \\
                 & $\rightarrow 3d \ ^3D_{3/2}$  &$1.968$     &$2.002(20)$     &$4.008(80)$                  \\

$9p \ ^2P_{1/2}$ & $\rightarrow 3d \ ^2D_{3/2}$  &$0.246$     &$0.250(4)$      &$0.062(2)$                  \\
                 & $\rightarrow 4s \ ^2S_{1/2}$  &$0.026$     &$0.016(3)$      &$0.0002(1)$                   \\
                 
$9p \ ^2P_{3/2}$ & $\rightarrow 3d \ ^2D_{5/2}$  &$0.334$     &$0.339(5)$      &$0.115(3)$                  \\
                 & $\rightarrow 3d \ ^2D_{3/2}$  &$0.111$     &$0.113(3)$      &$0.013(1)$                   \\
                 & $\rightarrow 4s \ ^2S_{1/2}$  &$0.041$     &$0.027(3)$      &$0.0010(2)$             \\

$8d \ ^2D_{5/2}$ & $\rightarrow 4p \ ^2P_{3/2}$  &$0.096$     &$0.286(5)$      &$0.082(3)$                     \\
                 
$8d \ ^2D_{3/2}$ & $\rightarrow 4p \ ^2P_{3/2}$  &$0.033$     &$0.101(3)$      &$0.010(1)$                     \\
                 & $\rightarrow 4p \ ^2P_{1/2}$  &$0.071$     &$0.221(4)$      &$0.049(2)$                  \\

$10s \ ^2S_{1/2}$& $\rightarrow 4p \ ^2P_{3/2}$  &$0.258$     &$0.242(5)$      &$0.058(2)$                     \\
                 & $\rightarrow 4p \ ^2P_{1/2}$  &$0.182$     &$0.171(3)$      &$0.029(1)$                  \\

\end{longtable}                                                              

%\end{center}      

\subsection{Electron attachment energies}
In Table \ref{tab1}, we present EAs corresponding to many states using our DHF, CCSD and
CCSD(T) methods and they are compared with the corresponding values given by National Institute of
Science and Technology (NIST) \cite{Nist}. We consider results obtained from the CCSD(T) method are
the final calculated results. In the above table, we also present differences between results
from the CCSD(T) method and quoted by NIST in percentage as $\Delta$. As seen in the table, all the 
calculated results are sub-one percent accurate. Among all other states, results of the $D$
states have large differences with the NIST values. Also, there are large differences between
the results obtained using DHF and CCSD(T) methods which indicate the amount of correlation 
effects involved to determine them. The differences between the results obtained from the CCSD and 
CCSD(T) methods are small implying contributions from the triple excitations are not very much 
significant in these calculations. Although the calculated EAs seem to be promising for their
sub-one percent accuracies, but we consider the experimental energies wherever required to determine other physical
properties in order to minimize uncertainties in the estimated results.

\begin{longtable*}{lcccccccc}
\caption{Wavelengths ($\lambda$ in \AA), transition rates ($A$ in $s^{-1}$), oscillator strengths ($f$) and branching ratios ($\Gamma$) from different works.
The values given in square bracket represent power of 10.} \label{tab3}\\
\hline \hline
Upper & Lower & $\lambda_{k i}$ & \multicolumn{2}{c}{$A_{k i}$} &  & \multicolumn{2}{c}{$f_{k i}$} & \multicolumn{1}{c}{$\Gamma_{k i}$}\\
\cline{4-5} \cline{7-8} state ($f$) &  state ($i$) &  & Others & Present & & Others & Present &  Present\\
\hline
\endfirsthead

\multicolumn{8}{c}{\tablename\ \thetable{} -- continued from previous page.} \\
\hline \hline
Upper & Lower & $\lambda_{k i}$ & \multicolumn{2}{c}{$A_{k i}$} &  & \multicolumn{2}{c}{$f_{k i}$} & $\Gamma_{k i}$\\
\cline{4-5} \cline{7-8}  state ($f$) &  state ($i$) &  & Others & Present & & Others & Present & Present\\
\hline \\
\endhead

\hline
\multicolumn{8}{c}{ {\it Continue} \dots } \\
\endfoot

\hline \hline 
\endlastfoot 

$4p_{1/2}$ & $\rightarrow 4s_{1/2}$ &  7701.08  &              & 3.785[7] & &2.93[-1]$^a$                 & 3.34[-1]   &1.0     \\

$4p_{3/2}$ & $\rightarrow 4s_{1/2}$ &  7666.99  &              & 3.834[7] & &5.88[-1]$^a$                 & 6.72[-1]   & $\sim$1.0     \\

$5s_{1/2}$ & $\rightarrow 4p_{3/2}$ &  12525.58 &1.58[7]$^b$   & 1.573[7] & &1.72[-1]$^a$, 1.86[-1]$^b$   & 1.84[-1]   &0.665     \\
           & $\rightarrow 4p_{1/2}$ &  12435.70 &7.95[6]$^b$   & 7.914[6] & &1.71[-1]$^a$, 1.84[-1]$^b$   & 1.82[-1]   &0.335              \\

$3d_{5/2}$ & $\rightarrow 4p_{3/2}$ &  11776.10  &2.38[7]$^b$  & 2.390[7] & &7.69[-1]$^a$, 7.42[-1]$^b$   & 7.40[-1]   & $\sim $ 1.0     \\
                 
$3d_{3/2}$ & $\rightarrow 4p_{3/2}$ &  11772.89  &3.97[6]$^b$  & 3.985[6] & &8.54[-1]$^a$, 8.24[-1]$^b$   & 8.20[-1]   &0.165     \\
           & $\rightarrow 4p_{1/2}$ &  11693.44  &2.01[7]$^b$  & 2.021[7] & &8.55[-1]$^a$, 8.25[-1]$^b$   & 8.24[-1]   &0.835              \\

$5p_{1/2}$ & $\rightarrow 3d_{3/2}$ &  31601.63  &1.65[6]$^b$  & 1.700[6] & &1.31[-1]$^a$, 1.23[-1]$^b$   & 1.26[-1]   &0.228     \\
           & $\rightarrow 5s_{1/2}$ &  27212.13  &             & 4.535[6] & &4.87[-1]$^a$                 & 4.99[-1]   &0.608             \\ 
           & $\rightarrow 4s_{1/2}$ &  4048.36   &             & 1.214[6] & &2.48[-3]$^a$                 & 2.96[-3]   &0.163             \\

$5p_{3/2}$ & $\rightarrow 3d_{3/2}$ &  31392.64  &1.66[5]$^b$  & 1.717[5] & &2.46[-2]$^b$                 & 2.52[-2]   &0.022             \\
           & $\rightarrow 3d_{5/2}$ &  31415.32  &1.50[6]$^b$  & 1.550[6] & &1.48[-1]$^b$                 & 1.52[-1]   &0.203              \\ 
           & $\rightarrow 5s_{1/2}$ &  27073.93  &             & 4.582[6] & &9.76[-1]$^a$                 & 1.001      &0.601              \\
           & $\rightarrow 4s_{1/2}$ &  4045.28   &             & 1.324[6] & &5.40[-3]$^a$                 & 6.48[-3]   &0.174              \\

$4d_{5/2}$ & $\rightarrow 5p_{3/2}$ &  37356.22  &             & 3.380[6] & &1.09$^a$                     & 1.054      &0.981         \\
           & $\rightarrow 4p_{3/2}$ &  6966.61   &1.37[4]$^b$  & 6.751[4] & &2.56[-4]$^a$, 1.49[-4]$^b$   & 7.324[-4]  &0.019               \\

$4d_{3/2}$ & $\rightarrow 5p_{3/2}$ &  37341.30  &             & 5.638[5] & &1.22[-1]$^a$                 & 1.17[-1]   &0.161          \\
           & $\rightarrow 5p_{1/2}$ &  37081.53  &             & 2.860[6] & &1.22[-1]$^a$                 & 1.17[-1]   &0.815              \\ 
           & $\rightarrow 4p_{3/2}$ &  6966.09   &2.40[3]$^b$  & 1.160[4] & &2.96[-5]$^a$, 1.75[-5]$^b$   & 8.39[-5]   &0.003              \\
           & $\rightarrow 4p_{1/2}$ &  6938.20   &1.90[4]$^b$  & 7.340[4] & &4.19[-4]$^a$, 2.75[-4]$^b$   & 1.05[-3]   &0.021          \\

$6s_{1/2}$ & $\rightarrow 5p_{3/2}$ &  36622.39  &             & 3.207[6] & &3.17[-1]$^a$                 & 3.20[-1]   &0.259         \\
           & $\rightarrow 5p_{1/2}$ &  36372.50  &             & 1.615[6] & &3.15[-1]$^a$                 & 3.18[-1]   &0.131              \\ 
           & $\rightarrow 4p_{3/2}$ &  6940.68   &4.95[6]$^b$  & 5.019[6] & &1.79[-2]$^b$                 & 1.80[-2]   &0.406              \\
           & $\rightarrow 4p_{1/2}$ &  6912.99   &2.50[6]$^b$  & 2.534[6] & &1.71[-2]$^a$, 1.79[-2]$^b$   & 1.80[-2]   &0.205          \\

$4f_{5/2}$ & $\rightarrow 4d_{3/2}$ &  137040.74 &             & 8.457[4] & &3.70[-1]$^a$                 & 3.54[-1]   &0.005     \\
           & $\rightarrow 4d_{5/2}$ &  136841.40 &             & 6.063[3] & &1.76[-2]$^a$                 & 1.69[-2]   &0.0004              \\ 
           & $\rightarrow 3d_{3/2}$ &  15172.52  &1.45[7]$^b$  & 1.490[7] & &8.24[-1]$^a$, 7.51[-1]$^b$   & 7.66[-1]   &0.928              \\
           & $\rightarrow 3d_{5/2}$ &  15167.21  &1.04[6]$^b$  & 1.065[6] & &3.92[-2]$^a$, 3.57[-2]$^b$   & 3.65[-2]   &0.066         \\

$4f_{7/2}$ & $\rightarrow 4d_{5/2}$ &  136840.10 &             & 9.097[4] & &3.53[-1]$^a$                 & 3.38[-1]   &0.006      \\
           & $\rightarrow 3d_{5/2}$ &  15167.21  &1.55[7]$^b$  & 1.599[7] & &7.85[-1]$^a$, 7.15[-1]$^b$   & 7.31[-1]   &0.994              \\

$6p_{1/2}$ & $\rightarrow 6s_{1/2}$ &  64576.12  &             & 1.038[6] & &6.41[-1]$^a$                 & 6.45[-1]   &0.345     \\
           & $\rightarrow 4d_{3/2}$ &  62456.16  &             & 9.061[5] & &2.61[-1]$^a$                 & 2.63[-1]   &0.301              \\ 
           & $\rightarrow 3d_{3/2}$ &  13400.73  &4.47[5]$^b$  & 4.527[5] & &6.48[-3]$^a$, 6.02[-3]$^b$   & 6.06[-3]   &0.151              \\
           & $\rightarrow 5s_{1/2}$ &  12542.77  &             & 4.214[5] & &9.42[-3]$^a$                 & 9.88[-3]   &0.140          \\
           & $\rightarrow 4s_{1/2}$ &  3448.36   &             & 1.871[5] & &2.51[-4]$^a$                 & 3.31[-4]   &0.062            \\

$6p_{3/2}$ & $\rightarrow 6s _{1/2}$&  64226.09  &             & 1.051[6] & &1.28$^a$                     & 1.292      &0.341     \\
           & $\rightarrow 4d _{3/2}$&  62128.66  &             & 9.125[4] & &5.21[-2]$^a$                 & 5.25[-2]   &0.030              \\ 
           & $\rightarrow 4d _{5/2}$&  62087.39  &             & 8.242[5] & &3.12[-1]$^a$                 & 3.16[-1]   &0.267             \\
           & $\rightarrow 3d _{3/2}$&  13385.59  &4.54[4]$^b$  & 4.606[4] & &1.31[-3]$^a$, 1.22[-3]$^b$   & 1.23[-3]   &0.149          \\
           & $\rightarrow 3d _{5/2}$&  13381.45  &4.54[4]$^b$  & 4.131[5] & &7.88[-3]$^a$, 7.36[-3]$^b$   & 7.35[-3]   &0.134           \\
           & $\rightarrow 5s _{1/2}$&  12529.51  &             & 4.433[5] & &1.98[-2]$^a$                 & 2.07[-2]   &0.144                 \\
           & $\rightarrow 4s _{1/2}$&  3447.36   &             & 2.154[5] & &5.91[-4]$^a$                 & 7.63[-4]   &0.070         \\

$5d_{5/2}$ & $\rightarrow 6p_{3/2}$ &  84923.52  &             & 8.267[5] & &1.380$^a$                    & 1.332      &0.590     \\
           & $\rightarrow 4f_{7/2}$ &  48605.27  &             & 2.085[5] & &6.10[-2]$^a$                 & 5.50[-2]   &0.150              \\ 
           & $\rightarrow 4f_{5/2}$ &  48605.27  &             & 1.043[4] & &4.07[-3]$^a$                 & 3.67[-3]   &0.007              \\
           & $\rightarrow 5p_{3/2}$ &  18297.93  &             & 1.176[5] & &6.10[-3]$^a$                 & 8.80[-3]   &0.084          \\
           & $\rightarrow 4p_{3/2}$ &  5833.51   &3.71[5]$^b$  & 2.379[5] & &2.47[-3]$^a$, 2.84[-3]$^b$   & 1.81[-3]   &0.170           \\

$5d_{3/2}$ & $\rightarrow 6p_{3/2}$ &  84886.93  &             & 1.379[5] & &1.53[-1]$^a$                 & 1.48[-1]   &0.098     \\
           & $\rightarrow 6p_{1/2}$ &  84283.10  &             & 6.999[5] & &1.530$^a$                    & 1.482      &0.510             \\ 
           & $\rightarrow 4f_{5/2}$ &  48593.28  &             & 2.184[5] & &5.69[-2]$^a$                 & 5.12[-2]   &0.156             \\
           & $\rightarrow 5p_{3/2}$ &  18296.22  &             & 1.986[4] & &6.83[-4]$^a$                 & 9.90[-4]   &0.014          \\
           & $\rightarrow 5p_{1/2}$ &  18233.65  &             & 1.082[5] & &7.49[-3]$^a$                 & 1.07[-2]   &0.077           \\
           & $\rightarrow 4p_{3/2}$ &  5833.33   &6.13[4]$^b$  & 3.924[4] & &2.72[-4]$^a$, 3.13[-4]$^b$   & 1.99[-4]   &0.028                 \\
           & $\rightarrow 4p_{1/2}$ &  5813.76   &2.86[5]$^b$  & 1.796[5] & &2.52[-3]$^a$, 2.90[-3]$^b$   & 1.81[-3]   &0.128         \\

$7s_{1/2}$ & $\rightarrow 6p _{3/2}$&  78955.34  &             & 9.807[5] & &4.56[-1]$^a$                 & 4.55[-1]   &0.145     \\
           & $\rightarrow 6p _{1/2}$&  78432.68  &             & 4.942[5] & &4.54[-1]$^a$                 & 4.53[-1]   &0.073              \\ 
           & $\rightarrow 5p _{3/2}$&  18004.69  &             & 1.140[6] & &2.70[-2]$^a$                 & 2.75[-2]   &0.168              \\
           & $\rightarrow 5p _{1/2}$&  17944.11  &             & 5.770[5] & &2.71[-2]$^a$                 & 2.77[-2]   &0.085          \\
           & $\rightarrow 4p _{3/2}$&  5803.37   &2.35[6]$^b$  & 2.375[6] & &5.68[-3]$^a$, 5.93[-3]$^b$   & 5.96[-3]   &0.351           \\
           & $\rightarrow 4p _{1/2}$&  5784.00   &1.19[6]$^b$  & 1.201[6] & &5.71[-3]$^a$, 5.95[-3]$^b$   & 5.99[-3]   &0.177                 \\

$5f_{7/2}$ & $\rightarrow 5d_{5/2}$ &  237255.75 &             & 6.019[4] & &6.87[-1]$^a$                 & 6.73[-1]   &0.007     \\
           & $\rightarrow 4d_{5/2}$ &  31156.02  &             & 2.101[6] & &3.93[-1]$^a$                 & 4.05[-1]   &0.239              \\ 
           & $\rightarrow 3d_{5/2}$ &  11022.87  &6.54[6]$^b$  & 6.612[6] & &1.76[-1]$^a$, 1.59[-1]$^b$   & 1.60[-1]   &0.754             \\

$5f_{5/2}$ & $\rightarrow 5d_{3/2}$ &  237539.62 &             & 5.593[4] & &7.20[-1]$^a$                 & 7.05[-1]   &0.006     \\
           & $\rightarrow 5d_{5/2}$ &  237255.75 &             & 4.012[3] & &3.43[-2]$^a$                 & 3.36[-2]   &0.0004             \\ 
           & $\rightarrow 4d_{3/2}$ &  31166.41  &             & 1.954[6] & &4.13[-1]$^a$                 & 4.24[-1]   &0.223            \\
           & $\rightarrow 4d_{5/2}$ &  31156.02  &             & 1.400[5] & &1.97[-2]$^a$                 & 2.02[-2]   &0.016        \\
           & $\rightarrow 3d_{3/2}$ &  11025.67  &6.10[6]$^b$  & 6.171[6] & &1.84[-1]$^a$, 1.67[-1]$^b$   & 1.68[-1]   &0.734           \\
           & $\rightarrow 3d_{5/2}$ &  11022.87  &4.36[5]$^b$  & 4.406[5] & &8.78[-3]$^a$, 7.94[-3]$^b$   & 7.98[-3]   &0.050                 \\

$5g_{7/2}$ & $\rightarrow 5f_{5/2}$ &  9451795.84&             & 0.524    & &                             & 9.30[-3]   &$\sim 0$     \\
           & $\rightarrow 5f_{7/2}$ &  9451795.84&             & 0.020    & &                             & 2.58[-4]   &$\sim 0$               \\ 
           & $\rightarrow 4f_{7/2}$ &  40169.35  &             & 1.532[5] & &3.70[-2]$^a$                 & 3.68[-2]   &0.036             \\
           & $\rightarrow 4f_{5/2}$ &  40169.35  &             & 4.136[6] & &1.340$^a$                    & 1.326      &0.964          \\

$5g_{9/2}$ & $\rightarrow 5f_{7/2}$ &  9451795.84&             & 0.543    & &                             & 9.04[-3]   &$\sim 0$     \\
           & $\rightarrow 4f_{7/2}$ &  40169.35  &             & 4.289[6] & & 1.300$^a$                    & 1.289      &1.000              \\

$7p_{1/2}$ & $\rightarrow 7s_{1/2}$ &  125683.20 &             & 3.343[5] & &7.87[-1]$^a$                 & 7.87[-1]   &0.211     \\
           & $\rightarrow 5d_{3/2}$ &  113102.67 &             & 4.187[5] & &3.88[-1]$^a$                 & 3.99[-1]   &0.265              \\ 
           & $\rightarrow 6s_{1/2}$ &  27630.49  &             & 1.446[5] & &1.60[-2]$^a$                 & 1.64[-2]   &0.091              \\
           & $\rightarrow 4d_{3/2}$ &  27234.95  &             & 2.946[5] & &1.70[-2]$^a$                 & 1.63[-2]   &0.186          \\
           & $\rightarrow 3d_{3/2}$ &  10489.97  &2.17[5]$^b$  & 2.194[5] & &1.93[-3]$^a$, 1.79[-3]$^b$   & 1.80[-3]   &0.139           \\
           & $\rightarrow 5s_{1/2}$ &  9956.84   &             & 1.193[5] & &1.66[-3]$^a$                 & 1.76[-3]   &0.075                 \\
           & $\rightarrow 4s_{1/2}$ &  3218.55   &             & 5.108[4] & &5.37[-5]$^a$                 & 7.88[-5]   &0.003          \\

$7p_{3/2}$ & $\rightarrow 7s _{1/2}$&  124976.36 &            & 3.383[5] & &1.570$^a$                     & 1.575      &0.201     \\
           & $\rightarrow 5d _{3/2}$&  112529.94 &            & 4.212[4] & &7.74[-2]$^a$                  & 7.95[-2]   &0.025              \\ 
           & $\rightarrow 5d _{5/2}$&  112466.19 &            & 3.805[5] & &4.64[-1]$^a$                  & 4.78[-1]   &0.227             \\
           & $\rightarrow 6s _{1/2}$&  27596.18  &            & 1.508[5] & &3.30[-2]$^a$                  & 3.42[-2]   &0.090          \\
           & $\rightarrow 4d _{3/2}$&  27201.61  &            & 2.984[4] & &3.44[-3]$^a$                  & 3.29[-3]   &0.018           \\
           & $\rightarrow 4d _{5/2}$&  27193.69  &            & 2.675[5] & &2.06[-2]$^a$                  & 1.96[-2]   &0.159                 \\
           & $\rightarrow 3d _{3/2}$&  10485.02  &2.21[4]$^b$ & 2.225[4] & &3.92[-4]$^a$, 3.64[-4]$^b$    & 3.64[-4]   &0.013          \\
           & $\rightarrow 3d _{5/2}$&  10482.49  &1.99[4]$^b$ & 2.580[5] & &2.35[-3]$^a$, 2.18[-3]$^b$    & 2.19[-3]   &0.154              \\
           & $\rightarrow 5s _{1/2}$&  9952.38   &            & 1.279[5] & &3.57[-3]$^a$                  & 3.78[-3]   &0.076             \\
           & $\rightarrow 4s _{1/2}$&  3218.08   &            & 6.187[4] & &1.36[-4]$^a$                  & 1.92[-4]   &0.037              \\

$6d_{5/2}$ & $\rightarrow 7p_{3/2}$ &  160990.92 &            & 2.755[5] & &                              & 1.594      &0.293     \\
           & $\rightarrow 5f_{5/2}$ &  91812.99  &            & 7.227[3] & &                              & 9.08[-3]   &0.008              \\ 
           & $\rightarrow 5f_{7/2}$ &  91812.99  &            & 1.446[5] & &                              & 1.36[-1]   &0.154              \\
           & $\rightarrow 6p_{3/2}$ &  37199.74  &            & 5.876[4] & &                              & 1.82[-2]   &0.063          \\
           & $\rightarrow 4f_{5/2}$ &  28026.51  &            & 4.489[3] & &                              & 5.25[-4]   &0.005            \\
           & $\rightarrow 4f_{7/2}$ &  28026.51  &            & 8.991[4] & &                              & 7.89[-3]   &0.096                 \\
           & $\rightarrow 5p_{3/2}$ &  14335.35  &            & 2.321[2] & &                              & 1.07[-5]   &0.0002          \\
           & $\rightarrow 4p_{3/2}$ &  5361.07   &4.86[5]$^b$ & 3.577[5] & &2.81[-3]$^a$, 3.14[-3]$^b$    & 2.30[-3]   &0.004              \\

$6d_{3/2}$ & $\rightarrow 7p_{3/2}$ &  160832.18 &            & 4.597[4] & &                              & 1.77[-1]   &0.049        \\
           & $\rightarrow 7p_{1/2}$ &  159676.53 &            & 2.333[5] & &                              & 1.773      &0.250             \\ 
           & $\rightarrow 5f_{5/2}$ &  91790.61  &            & 1.517[5] & &                              & 1.27[-1]   &0.162              \\
           & $\rightarrow 6p_{3/2}$ &  37196.07  &            & 9.882[3] & &                              & 2.04[-3]   &0.011          \\
           & $\rightarrow 6p_{1/2}$ &  37079.66  &            & 5.274[4] & &                              & 2.16[-2]   &0.056           \\
           & $\rightarrow 4f_{5/2}$ &  28024.42  &            & 9.475[4] & &                              & 7.40[-3]   &0.101                 \\
           & $\rightarrow 5p_{3/2}$ &  14334.81  &            & 38.69    & &                              & 1.18[-6]   &$\sim 0$            \\
           & $\rightarrow 5p_{1/2}$ &  14296.37  &            & 6.034[2] & &                              & 3.67[-5]   &0.0006                 \\
           & $\rightarrow 4p_{3/2}$ &  5360.99   &8.06[4]$^b$ & 5.991[4] & &3.10[-4]$^a$, 3.47[-4]$^b$    & 2.56[-4]   &0.064              \\
           & $\rightarrow 4p_{1/2}$ &  5344.45   &3.86[5]$^b$ & 2.849[5] & &2.96[-3]$^a$, 3.30[-3] $^b$   & 2.42[-3]   &0.305            \\

$8s_{1/2}$ & $\rightarrow 7p_{3/2}$ &  144722.68 &            & 3.779[5] & &5.93[-1]$^a$                  & 5.90[-1]   &0.093     \\
           & $\rightarrow 7p_{1/2}$ &  143786.27 &            & 1.905[5] & &5.91[-1]$^a$                  & 5.87[-1]   &0.047             \\ 
           & $\rightarrow 6p_{3/2}$ &  36262.54  &            & 3.760[5] & &3.60[-2]$^a$                  & 3.68[-2]   &0.093              \\
           & $\rightarrow 6p_{1/2}$ &  36151.89  &            & 1.906[5] & &3.60[-2]$^a$                  & 3.71[-2]   &0.047          \\
           & $\rightarrow 5p_{3/2}$ &  14193.98  &            & 6.079[5] & &8.77[-3]$^a$                  & 9.12[-3]   &0.150           \\
           & $\rightarrow 5p_{1/2}$ &  14156.31  &            & 3.009[5] & &8.84[-3]$^a$                  & 8.98[-3]   &0.074                 \\
           & $\rightarrow 4p_{3/2}$ &  5341.17   &1.31[6]$^b$ & 1.328[6] & &2.70[-3]$^a$, 2.81[-3]$^b$    & 2.82[-3]   &0.328         \\
           & $\rightarrow 4p_{1/2}$ &  532.481   &6.64[5]$^b$ & 6.700[5] & &2.71[-3]$^a$, 2.82[-3]$^b$    & 2.83[-3]   &0.166              \\ 

\end{longtable*} 
References: $^a$ \cite{Civis}; $^b$ \cite{safronova}
%\end{center}

\subsection{E1 matrix elements and oscillator strengths}
We present E1 matrix elements from the DHF and CCSD(T) methods along with their line strengths
from latter in Table \ref{tab2}. We estimate uncertainties associated with the calculations
using the CCSD(T) method as both DHF and CCSD methods are just part of this approach.
To estimate net uncertainty of an E1 matrix element, we take into account incompleteness
of basis functions and differences in the results obtained using the CCSD and CCSD(T) methods to estimate
uncertainty due to the approximation in the level of excitations. We also find that 
the correlation contributions, estimated
as the differences between the results obtained using the DHF and CCSD(T) methods, are significant in almost
all the transitions.

\begin{table}[h]
\caption{Lifetimes ($\tau$) of all the low-lying states up to $8S$ in K atom (in $ns$).} \label{tab4}
\begin{ruledtabular}
%\begin{center}
\begin{tabular}{lcccc}
State & This work & Others & Experiment \\ 
\hline \\
$5s \ ^2S_{1/2}$ & 42.31(1.1)   &46.50$^a$, 42.5$^b$                     \\

$3d \ ^2D_{5/2}$ & 41.85(1.2)   &45.85$^a$, 42.5$^b$       &                \\
$3d \ ^2D_{3/2}$ & 41.32(1.0)   &45.24$^a$, 41.9$^b$       &42(3)$^g$, 42(3)$^{d}$       \\

$5p \ ^2P_{1/2}$ & 134.40(3.0)  &127.05$^a$, 137.1$^b$     &137.6(1.3)$^k$            \\
                 &              &130$^h$                                               \\   
$5p \ ^2P_{3/2}$ & 131.10(3.0)  &124.02$^a$, 133.9$^b$     &133(3)$^f$, 134(2)$^{e}$       \\

$4d \ ^2D_{5/2}$ & 290.02(8.0)  &291.18$^a$, 293.9$^b$             \\
$4d \ ^2D_{3/2}$ & 284.10(7.3)  &285.56$^a$, 289.4$^b$              \\
  
$6s \ ^2S_{1/2}$ & 80.81(1.1)   &87.12$^a$, 81.4$^b$       &88(2)$^{d}$, 68(9)$^i$   \\ 
 
$4f \ ^2F_{5/2}$ & 62.30(2.0)   &70.65$^a$, 64.7$^b$                       \\
$4f \ ^2F_{7/2}$ & 62.20(2.0)   &70.65$^a$, 63.9$^b$                     \\
   
$6p \ ^2P_{1/2}$ & 332.70(8.2)  &321.67$^a$, 340.7$^b$     &344(3)$^{e}$                   \\
$6p \ ^2P_{3/2}$ & 324.20(7.1)  &312.77$^a$, 332.0$^b$     &333(3)$^{e}$, 310(15)$^f$                \\
    
$5d \ ^2D_{5/2}$ & 713.70(19.1) &769.63$^a$, 650.8$^b$                                              \\
$5d \ ^2D_{3/2}$ & 712.70(21.1) &767.41$^a$, 653.1$^b$     &572(14)$^{d}$, 610(90)$^j$       \\
 
$7s \ ^2S_{1/2}$ & 147.71(1.2)  &158.83$^a$, 148.8$^b$     &155(6)$^{d}$, 165(12)$^j$      \\
   
$5f \ ^2F_{7/2}$ & 114.11(3.2)  &125.70$^a$, 117.9$^b$                       \\
$5f \ ^2F_{5/2}$ & 114.12(3.1)  &125.70$^a$, 118.0$^b$     &117(3)$^c$               \\
                
$5g \ ^2G_{7/2}$ & 233.20(1.3)  &                                             \\
$5g \ ^2G_{9/2}$ & 233.20(1.3)  &                                                  \\

$7p \ ^2P_{1/2}$ &632.10(15.2)  &619.47$^a$, 648.6$^b$     &623(6)$^{e}$      \\
$7p \ ^2P_{3/2}$ &595.50(14.1)  &601.80$^a$, 632.0$^b$     &592(6)$^{e}$   \\
  
$6d \ ^2D_{5/2}$ &1066.10(24.1) &1168.54$^b$, 913.7$^a$                    \\
                 
$6d \ ^2D_{3/2}$ &1071.11(26.0) &1180.58$^a$, 925.7$^b$    &807(20)$^{d}$, 890(60)$^j$ \\
                 &              &1050$^h$                  &                              \\

$8s \ ^2S_{1/2}$ & 247.40(2.0)  &267.23$^a$, 250.7$^b$     &238(4)$^{d}$, 260(14)$^j$       \\  
          
\end{tabular}
%\end{center}
\end{ruledtabular}
References: $^a$ \cite{Theodosiou};$^b$ \cite{safronova} 
            $^c$ \cite{Glodz}; $^d$ \cite{Hart}; $^e$ \cite{Berends}; 
            $^f$\cite{Svanberg}; $^g$\cite{Teppner}; 
            $^h$\cite{Grudzev}; $^i$\cite{Happer};
            $^j$\cite{Gallagher}; $^k$\cite{Mills}
\end{table}

It is not always possible to obtain precise values of either E1 matrix elements
or transition strengths from the measured lifetimes due to association of many
transition probabilities with these quantities. However, the $4P$ states are the first 
two excited states which decay to the ground state only via one allowed transition each.
There are precise measurements of
lifetimes of these states are available in K atom. The lifetime of the $4P_{1/2}$
state is reported to be 26.69(5) $ns$ \cite{Wang}. By combining this result with the
experimental value of wavelength $\lambda=7701.1$ \AA \cite{Nist}
of the $4P_{1/2} \rightarrow 4S$ transition, we
find the E1 matrix element of the $4P_{1/2} \rightarrow 4S$ transition to be
4.110(5) a.u. against our calculated result 4.131(20) a.u. Similarly, the lifetime
of the $4P_{3/2}$ state is measured to be 26.34(5) $ns$ \cite{Wang}. This state
has an allowed transition channel to the ground state and it can also decay to the 
first excited $4P_{1/2}$ state via both the electric quadrupole (E2) and magnetic dipole
(M1) channels. It is found from our analysis that the transition probabilities of
an electron due to the above forbidden channels from the $4P_{3/2}$ state are very 
small and can be neglected within our estimated uncertainties. Therefore combining
the measured lifetime of the $4P_{3/2}$ state with the experimental value of the 
$4P_{3/2} \rightarrow 4S$ transition wavelength $\lambda =7667.0$ \AA, we obtain
E1 matrix element of the $4P_{3/2} \rightarrow 4S$ transition as 5.812(6) a.u. 
Although our calculated results from the CCSD(T) method are in agreement with the
extracted values from the experimental lifetimes of the $4P$ states, however
we consider below the precise values obtained from the measurements to evaluate
polarizabilities.

\begin{table}
\caption{\label{4sp}Scalar polarizability of the ground state in a.u. in K
along with individual contribution from various intermediate terms. Uncertainties
are quoted in the parentheses.}
\begin{ruledtabular}
\begin{tabular}{cccc}
\multicolumn{2}{c}{Contribution} &
\multicolumn{1}{c}{$\alpha^{(0)}_v$}\\
\hline
\multicolumn{2}{c}{$\alpha_v$} &  \\
$4s \rightarrow$ & $4p_{1/2}$ & $~ 95.165(231) $    \\
                       & $4p_{3/2}$ & $ 189.471(391)$      \\
                       & $5p_{1/2}$ & $~~~ 0.235(10)  $     \\
                       & $5p_{3/2}$ & $~~~ 0.514(10)  $     \\
                       & $6p_{1/2}$ & $~~~~ 0.019(3)  $    \\   
                       & $6p_{3/2}$ & $~~~~ 0.044(3)  $    \\   
                       & $7p_{1/2}$ & $~~~~ 0.004(3)  $    \\   
                       & $7p_{3/2}$ & $~~~~ 0.010(3)  $    \\   
                       & $8p_{1/2}$ & $~~~~ 0.001(3)  $    \\
                       & $8p_{3/2}$ & $~~~~ 0.003(2)  $    \\   
                       & $9p_{1/2}$ & $~~~~ 0.001(2)  $    \\   
                       & $9p_{3/2}$ & $~~~~ 0.002(2)  $     \\   
                       
\multicolumn{2}{c}{$\alpha_c$}                   & $~~5.131(0.30)  $  \\
\multicolumn{2}{c}{$\alpha_{cn}$}                & $-0.127(0.02)   $  \\
\multicolumn{2}{c}{$\alpha_{\rm{tail}}$}         & $~~0.062(0.01)  $  \\
\multicolumn{2}{c}{$\alpha_{\rm{total}}$}        & $~~~290.5(1.0)  $  \\
&& \\
\multicolumn{2}{c}{Experiment} &    292.8(6.1) \cite{molof}, 305.0(21.6)\cite{hall}  \\
                               & &  290.58(1.42)\cite{holmgren} \\
\multicolumn{2}{c}{Others} &        290.2(8)\cite{derevianko1}, 289.3 \cite{safronova} \\

\end{tabular}   
\end{ruledtabular}
\end{table}

For the astrophysical interest, we also evaluate transition probabilities, oscillator strengths
for emission lines and branching ratios taking all possible allowed transitions among the considered
states in K atom using the above E1 matrix elements and tabulated wavelengths in NIST \cite{Nist}.
These values are given in Table \ref{tab3}. Four decade ago, Anderson and Ziltis had carried out
calculations of these quantities using a Coulomb approximation in the non-relativistic method 
\cite{Anderson}. Safronova and Safronova have also reported few of these quantities using a 
linearized approximation to the CCSD method (SD method) \cite{safronova}. In a recent
paper, Civi$\check{\text{s}}$ et al have used a quantum defect theory to evaluate these 
quantities in many transitions which are of particular astrophysical interest and compared
with the estimated results from various measurements and observations \cite{Civis} including those
are given in \cite{Anderson}. In the above
table, we compare our results along with the values reported in \cite{safronova, Civis}. As seen
in the table, all the results are in fair agreement among each other. Many of these results
related to the $6D$ states were not known previously to compare with the present estimation. We also
give branching ratios explicitly of individual allowed transition in the same table by considering into 
account all possible allowed transitions and neglecting transition probabilities due to the
forbidden transitions.

\begin{table}
\caption{\label{4p1p}Scalar polarizability of the $4P_{1/2}$ state of K atom in a.u. Uncertainties are given in the parentheses.}
\begin{ruledtabular}
\begin{tabular}{cccc}
\multicolumn{2}{c}{Contribution} &
\multicolumn{1}{c}{$\alpha^{(0)}_v$}\\
\hline\\ 
\multicolumn{2}{c}{$\alpha_v$} &  \\
$4p_{1/2} \rightarrow$ & $4s_{1/2}$ &$-95.165(231)$    \\
                       & $5s_{1/2}$ &$136.679(705)$      \\
                       & $6s_{1/2}$ &$~~4.179(100)$     \\
                       & $7s_{1/2}$ &$~~~0.971(20)$     \\
                       & $8s_{1/2}$ &$~~~0.389(10)$    \\   
                       & $9s_{1/2}$ &$~~~0.188(10)$    \\   
                       & $10s_{1/2}$ &$~~~~0.003(5)$    \\   
                       & $3d_{3/2}$ &$ 545.86(683)$    \\   
                       & $4d_{1/2}$ &$~~~~0.246(8)$    \\
                       & $5d_{3/2}$ &$~~~0.296(10)$    \\   
                       & $6d_{3/2}$ &$~~~0.336(10)$    \\   
                       & $7d_{3/2}$ &$~~~~0.017(5)$     \\   
                       & $8d_{3/2}$ &$~~~~0.009(5)$    \\

\multicolumn{2}{c}{$\alpha_c$}             & $~5.131(0.30) $  \\
\multicolumn{2}{c}{$\alpha_{cn}$}          & $-0.0002(0.001)$    \\
\multicolumn{2}{c}{$\alpha_{\rm{tail}}$}   & $~~~~6.4(2.0)$  \\
\multicolumn{2}{c}{$\alpha_{\rm{total}}$}  & $~~~~~~606(7)$  \\
&& \\
\multicolumn{2}{c}{Experiment} &       587(87)  \cite{marrus}\\
\multicolumn{2}{c}{Others} &       604.1 \cite{safronova}, 602(11)\cite{arora1}, 697.4 \cite{merawa}  \\

\end{tabular}   
\end{ruledtabular}
\end{table}

\subsection{Lifetimes of few excited states}
As mentioned above, the lifetimes of the $4P$ states in K atom are measured
precisely by Wang et al \cite{Wang}. However, these quantities are not
measured for some of the low-lying states. In fact, many measured values
available for the states other than the $4P$ states are not very precise. Using our calculated 
transition probabilities given in Table \ref{tab3}, we determine 
lifetimes of all the considered states except for the $4P$ states. These
results are given in Table \ref{tab4} and compared with other theoretical
and experimental data in the same table. We have also estimated uncertainties to our estimated
lifetime values from the reported error bars of E1 matrix elements. Most of
our results are within the error bars of the experimental values. It can be
seen in the above table that the present values of lifetimes of many low-lying states 
match well with the other theoretical results given in \cite{safronova} and
differ from another calculated results reported in \cite{Theodosiou}; but
our results for some of the higher states match better with the values given in 
\cite{Theodosiou} than the results given in \cite{safronova}. Lifetimes
of the $5G$ states were not found elsewhere to compare with our results.

\begin{table}
\caption{\label{4p3p}Contributions to both the scalar and tensor polarizabilities
of the $4P_{3/2}$ state in K atom in a.u. Uncertainties are mentioned in the parentheses.}
\begin{ruledtabular}
\begin{tabular}{cccc}

\multicolumn{2}{c}{Contribution} &
\multicolumn{1}{c}{$\alpha^{(0)}_v$}&
\multicolumn{1}{c}{$\alpha^{(2)}_v$}\\
\hline\\ 
% & & \\
\multicolumn{2}{c}{$\alpha_v$} & \\
$4p_{3/2} \rightarrow$ & $4s_{1/2}$ &$~-95.56(21)$ &$~~95.156(21)$       \\   
                       & $5s_{1/2}$ & $138.81(101)$&$-138.81(101)$    \\
                       & $6s_{1/2}$ & $~~4.188(60)$&$-4.188(60)  $   \\
                       & $7s_{1/2}$ & $~~0.971(20)$&$-0.971(20)  $    \\
                       & $8s_{1/2}$ & $~~0.389(12)$&$-0.389(12)  $    \\
                       & $9s_{1/2}$ & $~~0.187(11)$&$-0.187(11)  $    \\
                       & $10s_{1/2}$& $~~0.106(11)$&$-0.106(11)  $    \\
                       & $3d_{3/2}$ & $~54.91(101)$&$~~43.93(74) $      \\
                       & $4d_{3/2}$ & $~~~0.021(3)$&$~~~0.016(3) $       \\
                       & $5d_{3/2}$ & $~~~0.033(4)$&$~~~0.026(4) $       \\
                       & $6d_{3/2}$ &$~~~0.036(4)$&$~~~0.029(4) $       \\
                       & $7d_{3/2}$ & $~~~0.026(4)$&$~~~0.021(4) $       \\
                       & $8d_{3/2}$ & $~~~0.018(3)$&$~~~0.015(3) $       \\
                       & $3d_{5/2}$ & $494.34(552)$&$-98.87(110) $   \\
                       & $4d_{5/2}$ & $~~~0.171(5)$&$~~-0.034(4) $   \\
                       & $5d_{5/2}$ &$~~~0.297(6)$&$~~-0.059(4) $   \\
                       & $6d_{5/2}$ & $~~~0.319(6)$&$~~-0.064(4) $   \\
                       & $7d_{5/2}$ & $~~~0.236(6)$&$~~-0.047(4) $   \\
                       & $8d_{5/2}$ & $~~~0.081(5)$&$~~-0.016(3) $  \\
 
\multicolumn{2}{c}{$\alpha_c$}                 & $5.131(0.30)$&$0.00$      \\
\multicolumn{2}{c}{$\alpha_{cn}$}              & $-0.0002(0.001)$&$0.0002(0.001)$    \\
\multicolumn{2}{c}{$\alpha_{\rm{tail}}$}       & $9.13(30)   $&$-1.4(0.2)   $      \\
\multicolumn{2}{c}{$\alpha_{\rm{total}}$}      & $614(6) $&$-106(2) $   \\
&& \\
\multicolumn{2}{c}{Experiment} &    613(103)\cite{marrus}, 614(10)\cite{krenn}&-107(2)\cite{krenn}   \\
\multicolumn{2}{c}{Others} &        614.1\cite{safronova}, 697.4\cite{merawa}, 635\cite{schmieder}  &
 $-107.9$\cite{safronova}, $-96$\cite{schmieder} \\
\end{tabular}   
\end{ruledtabular}
\end{table}

\subsection{Static dipole polarizabilities ($\omega = 0$) }
We evaluate static dipole polarizabilities of the ground, $4P$ and $3D$ states in
K atom using the reported E1 matrix elements and experimental energies of the most
important intermediate states. These results are reported in Tables \ref{4sp} to \ref{3d5p}
along with individual contributions explicitly from various intermediate states and compare 
them with other available experimental and calculations using both non-relativistic
and relativistic methods. As has been stated earlier, the core and core-valence correlation
contributions are estimated using the MBPT(3) method. Since these contributions are relatively  
smaller than the valence correlation contribution, the accuracies of the final results are
uninfluenced by these results. In fact, we also give estimated uncertainties to these results
by comparing E1 matrix elements obtained using this method with the CCSD(T) method. It can be noticed 
from these result tables that only few studies are carried out on polarizabilities in K.
Also, the experimental results from the direct measurements are not available precisely in any of
the considered states. Recently, there is a result on the ground state polarizability in K has been
reported by combining the measured ground state polarizability of sodium (Na) atom with the
ratio of these quantities between K and Na and reported as 290.58(1.42) a.u. \cite{holmgren}.
Similarly, the differential polarizability of the $4P_{1/2} \rightarrow 4S$ transition has been
measured to be 317.11(4) a.u. \cite{miller}. By combining this result with the above ground
state value, we refer the experimental value of the $4P_{1/2}$ state polarizability as 607.69(2.97) a.u. which
seem to be in excellent agreement with our calculated result. Other previous calculations reporting
this value in \cite{derevianko,safronova,arora1,merawa} are based on methods using lower 
approximations than the present work. There are no experimental results in the $3D$ states
available to compare the corresponding results with our calculations. However, a calculation
in the non-relativistic method using a pseudo-potential is reported on both the scalar and tensor
polarizabilities of the $3D$ states \cite{rerat}. These values are also given in Tables \ref{3d3p}
and \ref{3d5p}, which also seem to be fairly agreement with our calculations.

\begin{table}
\caption{\label{3d3p}Scalar and tensor polarizabilities of the $3D_{3/2}$ state
along with uncertainties in the parentheses given in a.u.}
\begin{ruledtabular}
\begin{tabular}{cccc}
\multicolumn{2}{c}{Contribution} &
\multicolumn{1}{c}{$\alpha^{(0)}_v$}&
\multicolumn{1}{c}{$\alpha^{(2)}_v$}\\
\hline\\ 
% & & \\
\multicolumn{2}{c}{$\alpha_v$} & \\
$3d_{3/2} \rightarrow$& $4p_{1/2}$ & $-272.93(343) $&$272.93(343)   $    \\ 
                      & $4p_{3/2}$ & $-55.29(103)  $&$~-44.23(75)   $    \\
                      & $5p_{1/2}$ & $612.30(234)  $&$-612.30(234)  $    \\
                      & $5p_{3/2}$ & $~120.8(121)  $&$~~~96.62(955) $    \\
                      & $6p_{1/2}$ & $~5.271(204)  $&$-5.271(204)   $    \\ 
                      & $6p_{3/2}$ & $~~1.068(20)  $&$~~~~0.854(10) $    \\ 
                      & $7p_{1/2}$ & $~~0.959(20)  $&$~~-0.959(20)  $    \\ 
                      & $7p_{3/2}$ & $~~~0.194(5)  $&$~~~~~~0.155(5)$    \\
                      & $8p_{1/2}$ & $~~0.352(20)  $&$~-0.352(20)   $     \\
                      & $8p_{3/2}$ & $~~~0.071(6)  $&$~~~~~~0.057(5)$    \\
                      & $9p_{1/2}$ & $~~0.200(10)  $&$~-0.200(10)   $    \\
                      & $9p_{3/2}$ & $~~~0.041(5)  $&$~~~~~~0.033(5)$    \\
                      & $4f_{5/2}$ & $~855.2(166)  $&$-171.03(332)  $    \\
                      & $5f_{5/2}$ & $~98.78(161)  $&$~-19.76(33)   $    \\
                      & $6f_{5/2}$ & $29.697(620)  $&$~~-5.94(13)   $    \\
                      & $7f_{5/2}$ & $13.058(260)  $&$~-2.611(53)   $     \\     

\multicolumn{2}{c}{$\alpha_c$}                & $5.131(0.30) $&$~~~~0.00     $         \\                                                                                                           
\multicolumn{2}{c}{$\alpha_{cn}$}             & $-0.03(0.01) $&$~~0.02(0.01) $           \\
\multicolumn{2}{c}{$\alpha_{\rm{tail}}$}      & $50.6(10.0)  $&$~-10.6(3.0)  $       \\
\multicolumn{2}{c}{$\alpha_{\rm{total}}$}     & $1465.5(21.5)$&$-502.6(12.5) $        \\
&& \\
\multicolumn{2}{c}{Others \cite{rerat}} &       1613   &  $-710$ \\
\end{tabular}   
\end{ruledtabular}
\end{table}

\begin{table}
\caption{\label{3d5p}Contributions to both the scalar and tensor polarizabilities of the
$3D_{5/2}$ state in K (in a.u.) with their uncertainties in the parentheses.}
\begin{ruledtabular}
\begin{tabular}{cccc}
\multicolumn{2}{c}{Contribution} &
\multicolumn{1}{c}{$\alpha^{(0)}_v$}&
\multicolumn{1}{c}{$\alpha^{(2)}_v$}\\
\hline\\ 
% & & \\
\multicolumn{2}{c}{$\alpha_v$} & \\
$3d_{5/2} \rightarrow$ & $4p_{3/2}$ & $-331.79(371) $& $331.79(371)  $    \\   
                       & $5p_{3/2}$ & $~~724.7(240) $& $-724.7(239)  $     \\
                       & $6p_{3/2}$ & $~~6.378(182) $& $-6.378(182)  $    \\ 
                       & $7p_{3/2}$ & $~~~1.168(20) $& $~-1.168(20)  $    \\ 
                       & $8p_{3/2}$ & $~~~0.425(15) $& $~-0.425(15)  $    \\ 
                       & $9p_{3/2}$ & $~~~0.246(12) $& $~-0.246(12)  $    \\
                       & $4f_{5/2}$ & $~~~40.7(101) $& $~~~46.54(112)$     \\
                       & $4f_{7/2}$ & $~~814.7(144) $& $-290.96(512) $    \\
                       & $5f_{5/2}$ & $~~4.698(140) $& $~~~5.369(160)$    \\
                       & $5f_{7/2}$ & $~~93.99(161) $& $~-33.57(58)  $    \\
                       & $6f_{5/2}$ & $~~~1.413(20) $& $~~~~1.615(20)$    \\
                       & $6f_{7/2}$ & $~~27.94(485) $& $~-9.98(173)  $     \\
                       & $7f_{5/2}$ & $~~~~0.178(5) $& $~~~~~0.203(5)$     \\
                       & $7f_{7/2}$ & $~~~12.36(23) $& $~~-4.42(12)  $      \\
                                                           
\multicolumn{2}{c}{$\alpha_c$}                   & $5.131(0.30) $ &$0.0            $     \\
\multicolumn{2}{c}{$\alpha_{cn}$}                & $-0.03(0.01) $ &$~~0.03(0.01)   $    \\
\multicolumn{2}{c}{$\alpha_{\rm{tail}}$}         & $50.6(10.0)  $ &$~ -15.4(4.0)   $      \\
\multicolumn{2}{c}{$\alpha_{\rm{total}}$}        & $1452.8(32.5)$ &$-701.7(25.6)   $      \\
&& \\
\multicolumn{2}{c}{Others} &       1613\cite{rerat}  &  $-710$\cite{rerat} \\
\end{tabular}   
\end{ruledtabular}
\end{table}

\begin{figure}[h]
  \includegraphics[scale=0.50, angle=270]{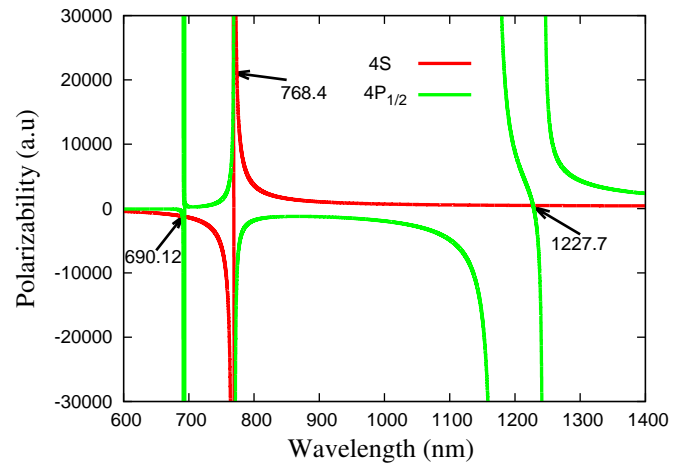}
  \caption{(color online) Dynamic polarizabilities (in a.u.) plots of the $4S$ (shown in red) and $4P_{1/2}$ 
(shown in green) states against wavelengths (in $nm$). Intersection points are identified as magic 
wavelengths of the $4P_{1/2}-4S$ transition. Resonance lines correspond at infinite values of 
polarizabilities.}
\label{fig1}
\end{figure}

\begin{figure}[h]
  \includegraphics[scale=0.50, angle=270]{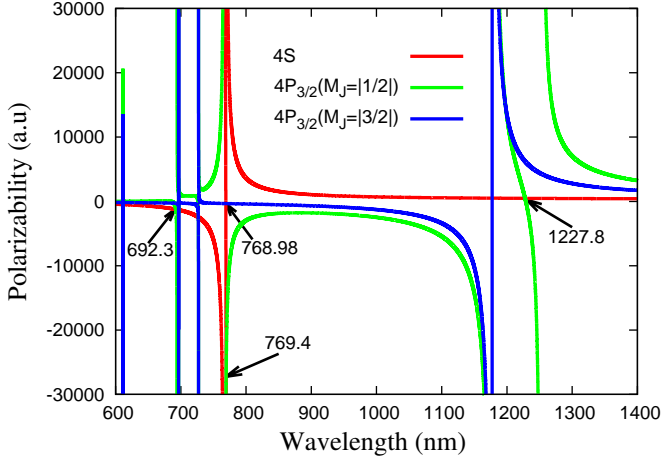}
  \caption{(color online) Dynamic polarizabilities (in a.u.) plots of the $4S$ and $4P_{3/2}$ 
states with different $M_J$ values against wavelengths (in $nm$). Lines shown with red, green and blue 
colors represent for results of the $4S$ state, $4P_{3/2}$ state with $M_J=1/2$ and $4P_{3/2}$ state 
with $M_J=3/2$, respectively. Magic wavelengths are shown by arrows.}
\label{fig2}
\end{figure}

\subsection{Reexamination of magic wavelengths}
Since our reported static polarizabilities for the ground and $4P$ states seem to be
more accurate than the results reported in \cite{arora1}, we intend here to further
study the dynamic polarizabilities and estimate the magic wavelengths in the
$4P_{1/2} \rightarrow 4S$ and $4P_{3/2} \rightarrow 4S$ transitions to compare them with the
values given in \cite{arora1}. In Fig. \ref{fig1},
we plot the dynamic polarizabilities of both the $4S$ and $4P_{1/2}$ states and look
for the intersections outside the resonance lines. From this plot, we give the magic
wavelengths corresponding to the intersection points of polarizabilities from both the 
states. The polarizability values and magic wavelengths around different resonance lines
for the $4P_{1/2} \rightarrow 4S$ transition are given in Table \ref{mgwv1}. From the analysis
of the above plot, we compare our results with the ones given in \cite{arora1}. We also find
same number of magic wavelengths as in \cite{arora1}, but the corresponding values
are further fine-tuned in this work due to more accurate values of the polarizabilities.
We also find similar results for the $4P_{3/2} \rightarrow 4S$ transition and they are
given in Table \ref{mgwv2}. As pointed out in \cite{arora1} and also found in the present
work, there is only one magic wavelength found for the $M_J=3/2$ sublevel of the
$4P_{3/2}$ state. It has to be noted that for the linearly polarized light, the Stark shift
of an energy level is independent of the sign of the $M_J$ value. However, this shift
depends on the sign of the $M_J$ value in the circularly polarized light due to the
presence of the vector component of the polarizability. It is found in our another recent
study in rubidium (Rb) atom \cite{arora2} that it is possible to obtain more number of magic wavelengths using
the circularly polarized light than the linearly polarized light and
we also anticipate for similar results in the $4P_{3/2} \rightarrow 4S$ transition in K
using the circularly polarized light.

\begin{table}
\caption{\label{mgwv1} Comparison of magic wavelengths and their corresponding polarizability
values around different resonance lines for the $4P_{1/2} \rightarrow 4S$ transition with 
other works.}
\begin{ruledtabular}
\begin{tabular}{ccccc}
\multicolumn{1}{c}{$\lambda_{res}$}&
\multicolumn{2}{c}{$\lambda_{magic}$}&
\multicolumn{2}{c}{$\alpha (\lambda_{magic})$}\\
\cline{2-3} \cline{4-5} & Present & Ref. \cite{arora1} & Present & Ref. \cite{arora1} \\
\hline\\ 

                        $1243.57$ & $1227.2(2)$ &1227.7(2) & 474(2)    &472(1)      \\
                                                &                                    \\

                        $1169.34$  &            &                                    \\
                        $770.11 $  &            &                                   \\
                                   &$768.412(3)$&768.413(4)& 21072(45) &20990(80)    \\  
                                   &            &                                  \\

                        $693.30$   &            &                                   \\
                        $691.12$   &            &                                  \\
                                   & $690.12(2)$&690.15(1) & $-1190(3)$&$-1186(2)$ \\   
\end{tabular}   
\end{ruledtabular}
\end{table}
               
\begin{table}
\caption{\label{mgwv2}Magic wavelengths and their corresponding dynamic polarizabilities
for the $4P_{3/2} \rightarrow 4S$ transition with $M_J=1/2$ and $M_J=3/2$ values from different
works.}
\begin{ruledtabular}
\begin{tabular}{ccccc}
\multicolumn{1}{c}{$\lambda_{res}$}&
\multicolumn{2}{c}{$\lambda_{magic}$}&
\multicolumn{2}{c}{$\alpha (\lambda_{magic})$}\\
\cline{2-3} \cline{4-5} & Present & Ref. \cite{arora1} & Present & Ref. \cite{arora1} \\
\hline\\ 
 $|M_{J_{3/2}}|=1/2$\\
&& \\
                        $1252.56$&$1227.8(2)$&1227.7(2) &$474(2)    $&472(1)     \\  
                                 &           &                                    \\    
                        $1177.61$&           &                                   \\
                        $1177.29$&           &                                   \\ 
                                 &$769.43(2)$&769.432(2)&$-27267(63)$&$-27190(60)$  \\    
                                 &           &                                   \\     
                        $766.70$ &           &                                    \\  
                        $696.66$ &           &                                  \\
                        $696.61$ &           &                                  \\
                        $694.07$ &           &                                   \\
                                 &$692.26(3)$&692.32(2) &$-1230(3)  $&$-1226(3)$   \\ 
 $|M_{J_{3/2}}|=3/2$\\
&& \\
                        $1177.61$&           &                                   \\
                        $1177.29$&           &                                   \\             
                                 &768.98(2)  &768.980(3)&$-336.52(6)$&$-356(8)$ \\
                       
\end{tabular}   
\end{ruledtabular}
\end{table}

\section{Conclusion}
We have studied electron attachment energies, electric dipole matrix elements, oscillator strengths, lifetimes
and dipole polarizabilities in potassium atom. Some of the reported results are improved significantly than
the previously known results. We affirm the reported magic wavelengths in the considered atom using the
dynamic polarizabilities. Our oscillator strength results will be very useful in the astrophysical studies
and other improved results will be very helpful in guiding the future experiments.

\section*{Acknowledgment}
We thank B. Arora for discussions regarding magic wavelengths. The calculations were carried out using PRL 3TFLOP HPC cluster, Ahmedabad.

\end{document}